# Why price inflation in developed countries is systematically underestimated

Ivan O Kitov

Institute for the Geopsheres' Dynamics, Russian Academy of Sciences

**Abstract**
There is an extensive historical dataset on real GDP per capita prepared by Angus Maddison. This dataset covers the period since 1870 with continuous annual estimates in developed countries. All time series for individual economies have a clear structural break between 1940 and 1950. The behavior before 1940 and after 1950 can be accurately ($R^2$ from 0.7 to 0.99) approximated by linear time trends. The corresponding slopes of regressions lines before and after the break differ by a factor of 4 (Switzerland) to 19 (Spain). We have extrapolated the early trends into the second interval and obtained much lower estimates of real GDP per capita in 2011: from 2.4 (Switzerland) to 5.0 (Japan) times smaller than the current levels. When the current linear trends are extrapolated into the past, they intercept the zero line between 1908 (Switzerland) and 1944 (Japan). There is likely an internal conflict between the estimating procedures before 1940 and after 1950. A reasonable explanation of the discrepancy is that the GDP deflator in developed countries has been highly underestimated since 1950. In the USA, the GDP deflator is underestimated by a factor of 1.4. This is exactly the ratio of the interest rate controlled by the Federal Reserve and the rate of inflation. Hence, the Federal Reserve actually retains its interest rate at the level of true price inflation when corrected for the bias in the GDP deflator.

**Introduction**

Real Gross Domestic Product (GDP) is not a directly measured macroeconomic variable. By definition, real GDP is estimated from actually measured nominal GDP (NGDP) which is corrected for the change in prices of all goods and services. The latter is defined as the GDP deflator, which is estimated and (somewhat subjectively) evaluated according to an extended set of complex procedures described in National Income and Product Accounts' documentation (BEA, 2011). Therefore, real GDP is a virtual value which depends on definitions of nominal GDP and GDP deflator. As a result, the overall behavior of real GDP should change together with the GDP deflator or NGDP. In this paper, we investigate the change in time trend observed in real GDP per capita and its consequences for the estimates of price inflation. In developed countries, this time trend has a clear break from 1940 to 1950 with the slope jumping by an order of magnitude.

For the purposes of our study, there are two important problems of the evolution of real GDP per capita in developed countries. The first problem is whether the time series are stationary since the start of measurements? It was formulated by Nelson and Plosser (1982) and has been in the centre of macroeconomic discussion on the permanent effect of shocks to output ever since (e.g. Ben-David *et al*., 2003; Narayan, 2006; Cuestas and Garratt, 2008). Stationarity was reported in some studies (e.g. Beechey and Österholm, 2008; Vougas, 2007). This is practically equivalent to a constant annual increment (Kitov, 2009; Kitov and Kitov, 2012).

In many studies, however, the null of a unit root in the output time series is not rejected (e.g. Michelacci and Zaffaroni, 2000; Mayoral, 2006). This is likely the result of structural breaks which cannot be ignored in unit root tests and present the second important aspect of our study. Chen (2008) used the Lagrange multiplier unit root test with two adopted structural breaks and rejected the null of a unit root in 11 from 19 developed countries including Australia, Japan, the US, the UK, and Germany. Caporale and Gil-Alana (2009) found a break in the second quarter of 1978 in the quarterly real per capita GDP series for the US and showed that, statistically, the best time trend is a linear one. This result supports our analysis of the trends and breaks in real per capita GDP time series (Kitov and Kitov, 2012).

In Section 1, we consider a model based on Maddison (2004) historical dataset with a (piece-wise) constant annual increment of real per capita GDP (mean-reverting around a linear deterministic trend) and a structural break. The break is artificial and likely associated with the change in measuring units/procedures. Linear regressions are calculated in both segments and extrapolated into the counterparts. In Section 2, by extrapolating the earlier trend into the second half of the1900s we re-estimate the real per capita GDP and thus the GDP deflator since 1950. For the US, the updated inflation series is compared to the interest rate defined by the Federal Reserve.

1. **Model and Maddison dataset**

Under our empirical framework (Kitov, 2009), real GDP per capita, $G_t$, in developed countries grows as a linear function of time (we call it inertial growth):

$$G_t = At + C \qquad (1)$$

Relationship (1) defines a linear trajectory of GDP per capita, where $C=G_{t0}$ and $t_0$ is the starting time. In the regime of inertial growth, the real GDP per capita increases by a constant value $A$ per time unit. The relative rate of growth along the inertial linear growth trend, $g_t$, is the reciprocal function of $G$:

$$g_t = A/G_t \qquad (2)$$

Relationship (2) implies that the rate of GDP growth will be asymptotically approaching zero (diminishing returns), but the annual increment $A$ will always be constant. This is different from the Solow-type model where the rate of growth is a positive (nonzero) value. In our model, the absolute rate of GDP growth is constant and is equal to *A* [$/y]. This constant annual increment thus defines the constant "speed" of economic growth in a one-to-one analogy with Newton's first law. Hence, one can consider the property of constant speed of real economic growth as "inertia of economic growth" or simply "inertia".

In Figure 1, we present real GDP per capita in fourteen biggest developed economies as a function of time. One can see two distinct periods of linear growth: before 1940 and after 1950. There is a ten-year period of very high turbulence induced by the Second World War. In essence, the global economy did not perform according to purely economic laws during this period. In addition, the concept of Gross Domestic Product was introduced and implemented during this short period. The measured GDP time series in developed countries are all started in 1950. The GDP values are reconstructed before 1950 from incomplete data. One might consider the period between 1940 and 1950 as a transition to new units and procedures of GDP measurements. In any case, there is a clear break in all time series somewhere between 1940 and 1950. Without loss of accuracy we exclude this period from our analysis.

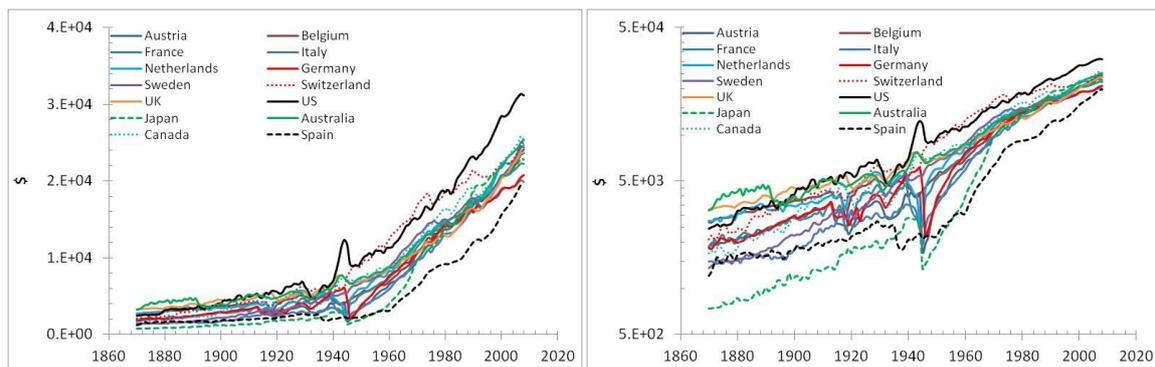

**Figure 1. The evolution of real GDP per capita (Maddison data set) from 1870 to 2008. The left panel presents all time series in the *lin-lin* scale and the right panel in the *lin-log* scale. Except the US and Canada, no country reveals exponential growth. There are rather two segments with sustainable linear trends and a break between 1940 and 1950.**

We have calculated linear regressions for all time series in both segments. Table 1 lists both estimated slopes in all countries and their ratio. Notice that we have replaced Maddison dataset with the Total Economy Database (TED) after 1950 (Conference Board, 2012). They are practically identical before 2008 and the TED includes estimates through 2011. (Only Japan and Spain show significant discrepancy between TED and Maddison estimates during the most recent period. For Japan, the TED and historical curve started to deviate in 1993.) The largest ration belongs to Spain (19.2) and the lowermost to Switzerland (4.0). This is a dramatic break in the relevant linear time trends. For the US, the ratio is 6.4. The split into two time segments is an alternative to exponential growth. Only the US and Canada demonstrate (likely by construction) a quasi-straight line in the *lin-log* coordinates, as shown in the right panel of

Figure 1. Since other countries are far from an exponential path during the whole period we prefer to divide it into two linear segments. This hypothesis is likely a superior one considering the differences in time series before and after the break.

We have already shown (Kitov and Kitov, 2012) that the first differences of real per capita GDP in these countries have no statistically significant trends before 1940 and after 1950. This is equivalent to linear time trends in the GDP series, as predicted by (1). In addition, Tables 2ab demonstrate that the first differences are stationary in both periods. (Kitov and Kitov (2012) showed that the fluctuations before 1940 are not normally distributed.) The Augmented Dickey-Fuller (ADF) and the DF-GLS tests with 4 time lags reject the null of unit roots in almost all cases except Spain and, in part, Japan. As mentioned above, these two countries have a degree of uncertainty in the GDP estimates and also had extended periods of severe political, social and economic turbulence in the past, and thus, might diverge from true economic growth. In the US GDP series between 1869 and 2001, Mayoral (2006) found nonstationarity (order of integration from 0.5 to 1.0), but she did not include a structural break. In any case, the length of the involved time series allows appropriate statistical power for all statistical tests. All in all, the annual increments are stationary and have no significant trends in both periods. It is possible to consider both periods separately and to extrapolate corresponding trends into their counterparts.

**Table 1. Slopes of the real GDP per capita curves between 1950 and 2011 (TED) and from 1870 and 1940 (Maddison).**

| Country | Slope, TED | Slope, Maddison | Ratio |
|---|---|---|---|
| **Australia** | 310.2 | 25.66 | 12.1 |
| **Austria** | 349.9 | 21.44 | 16.3 |
| **Belgium** | 321.3 | 32.26 | 10.0 |
| **Canada** | 314.0 | 48.82 | 6.4 |
| **France** | 298.1 | 38.19 | 7.8 |
| **Germany** | 339.9 | 36.26 | 9.4 |
| **Italy** | 348.2 | 24.64 | 14.1 |
| **Japan** | 286.2 | 30.45 | 9.4 |
| **Netherlands** | 319.5 | 38.90 | 8.2 |
| **Spain** | 277.3 | 14.46 | 19.2 |
| **Sweden** | 299.0 | 52.88 | 5.7 |
| **Switzerland** | 247.2 | 61.41 | 4.0 |
| **UK** | 282.7 | 39.44 | 7.2 |
| **US** | 387.7 | 60.86 | 6.4 |

In Figure 2, we extrapolate the TED time series back to 1870 and the Maddison historical estimates are extrapolated through 2011. All estimates are in 1990 International

Geary-Khamis dollars. As expected, the time series before 1940 and after 1950 are both well approximated by linear time trends. Interestingly, the extrapolated TED curves for Japan and Spain intersect the zero line in 1944 and 1943, respectively. For Switzerland, the extrapolated curve intersects the x-axis in 1908. For other countries, the intercept years vary between 1920 and 1940. This observation shows that a structural break in needed to avoid a negative GDP per capita before 1940. Such a structural break is likely of artificial nature and the historical estimates before 1940 might express the true (real) economic growth.

Table 2a. Unit root tests in the first difference between 1870 and 1940.

| Country | ADF | DF-GLS (lags) | | | |
|---|---|---|---|---|---|
| | | 4 | 3 | 2 | 1 |
| **Austria** | -5.27 | -3.056 | -2.928 | -3.637 | -4.837 |
| **Belgium** | -7.07 | -3.619 | -3.230 | -3.746 | -5.641 |
| **France** | -8.13 | -3.225 | -3.432 | -4.494 | -6.135 |
| **Germany** | -7.04 | -2.831 | -3.060 | -3.807 | -5.172 |
| **Italy** | -7.23 | -4.420 | -5.007 | -5.286 | -4.928 |
| **Netherlands** | -7.62 | -3.125 | -3.197 | -3.259 | -4.028 |
| **Spain** | -7.35 | -1.338 | -3.174 | -3.530 | -4.298 |
| **Sweden** | -7.25 | -3.060 | -2.726 | -3.489 | -4.932 |
| **Switzerland** | -6.87 | -2.878 | -3.168 | -4.187 | -5.399 |
| **UK** | -5.90 | -2.341 * | -3.399 | -3.652 | -3.737 |
| **US** | -7.30 | -4.457 | -4.488 | -4.446 | -5.493 |
| **Japan** | -8.52 | -1.038 * | -2.072 * | -3.220 | -5.868 |
| **Australia** | -8.35 | -3.839 | -3.736 | -3.710 | -4.063 |
| **Canada** | -5.53 | -3.653 | -4.307 | -4.253 | -4.465 |

*null rejected

Table 2b. Unit root tests in the first difference between 1950 and 2008.

| Country | ADF | DF-GLS (lags) | | | |
|---|---|---|---|---|---|
| | | 4 | 3 | 2 | 1 |
| **Austria** | -5.999 | -3.988 | -3.862 | -3.506 | -4.676 |
| **Belgium** | -6.819 | -3.209 | -3.100 | -3.672 | -4.002 |
| **France** | -5.36 | -2.916 | -3.508 | -3.363 | -4.170 |
| **Germany** | -7.894 | -1.725 | -2.186 | -2.382 | -4.065 |
| **Italy** | -5.856 | -2.669 | -3.855 | -3.978 | -4.988 |
| **Netherlands** | -4.605 | -2.463 | -3.302 | -3.542 | -3.605 |
| **Spain** | -3.136 | -1.724 * | -2.049 * | -2.234 * | -2.413 * |
| **Sweden** | -3.787 | -2.394 * | -2.948 | -3.105 | -3.821 |
| **Switzerland** | -5.455 | -3.985 | -3.684 | -4.276 | -5.407 |
| **UK** | -4.688 | -1.633 * | -2.111 * | -2.566 * | -3.453 |
| **US** | -5.811 | -3.046 | -3.689 | -4.436 | -4.928 |
| **Japan** | -4.529 | -1.965 * | -2.393 * | -2.226 * | -3.571 |
| **Australia** | -5.417 | -1.616 * | -2.527 * | -2.811 | -4.559 |
| **Canada** | -5.351 | -2.660 * | -3.317 | -3.705 | -4.132 |

There are important implications of the constant annual increment (diminishing returns) for economic policy in developed countries. Economists and economic authorities (e.g. the FRB and CBO) are waiting for a significant increase in the rate of real economic growth to close so called output gap, i.e. the difference between the measured level of real GDP and that expected form exponential extrapolation of the trend observed before 2007. In reality, there is no output gap, as Figure 2 demonstrates. Only Italy and Japan are far below the linear trend in real GDP per capita, and Australia is slightly above the expected level. France is also slightly below its long term linear trend. These countries might expect a mid-term recovery to the trend in the long run. For Australia, we expect a negative correction in the near future. The experience of Japan and Ireland clearly indicates that no positive deviation from the linear trend lasts long (Kitov, 2009).

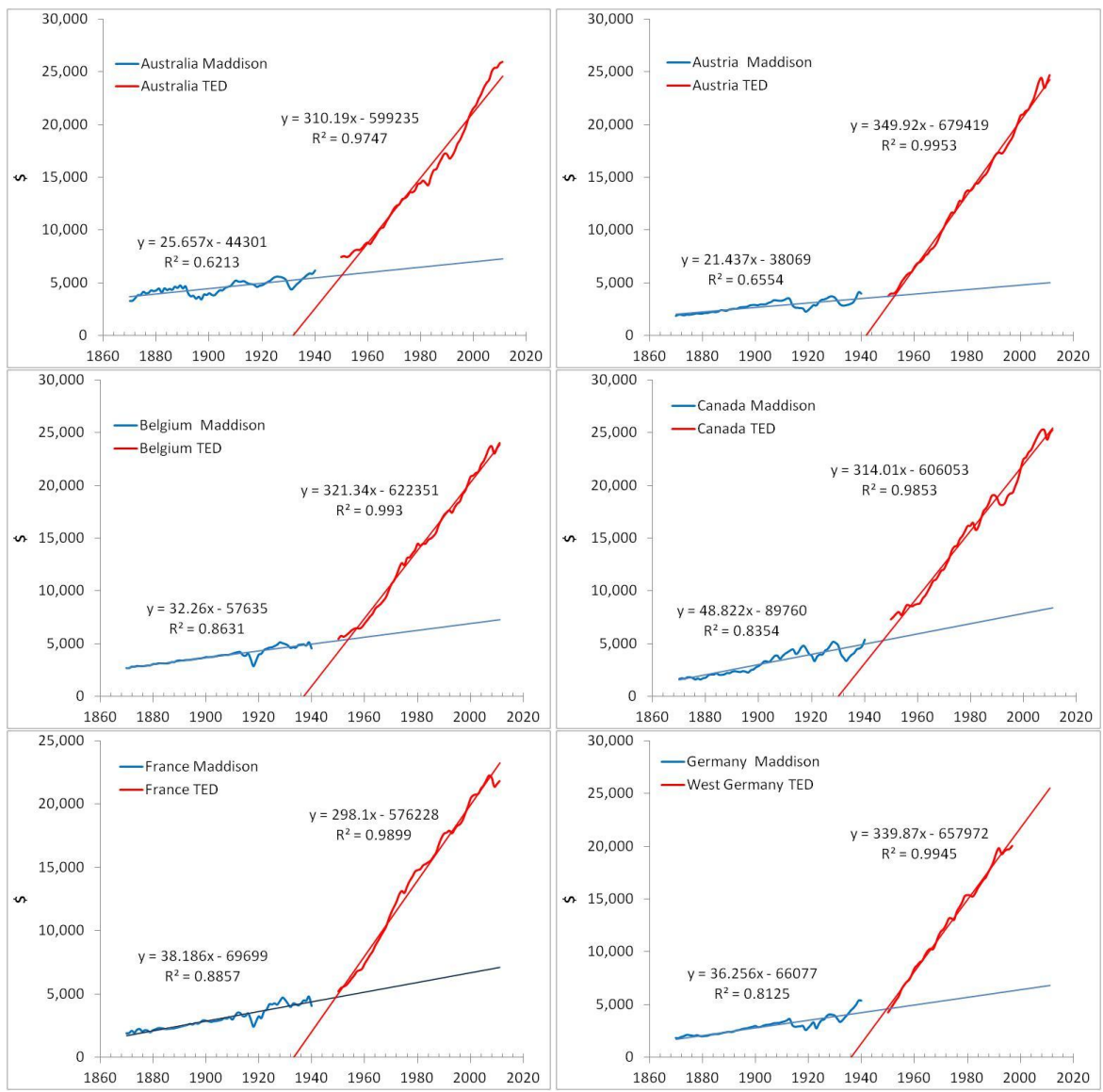

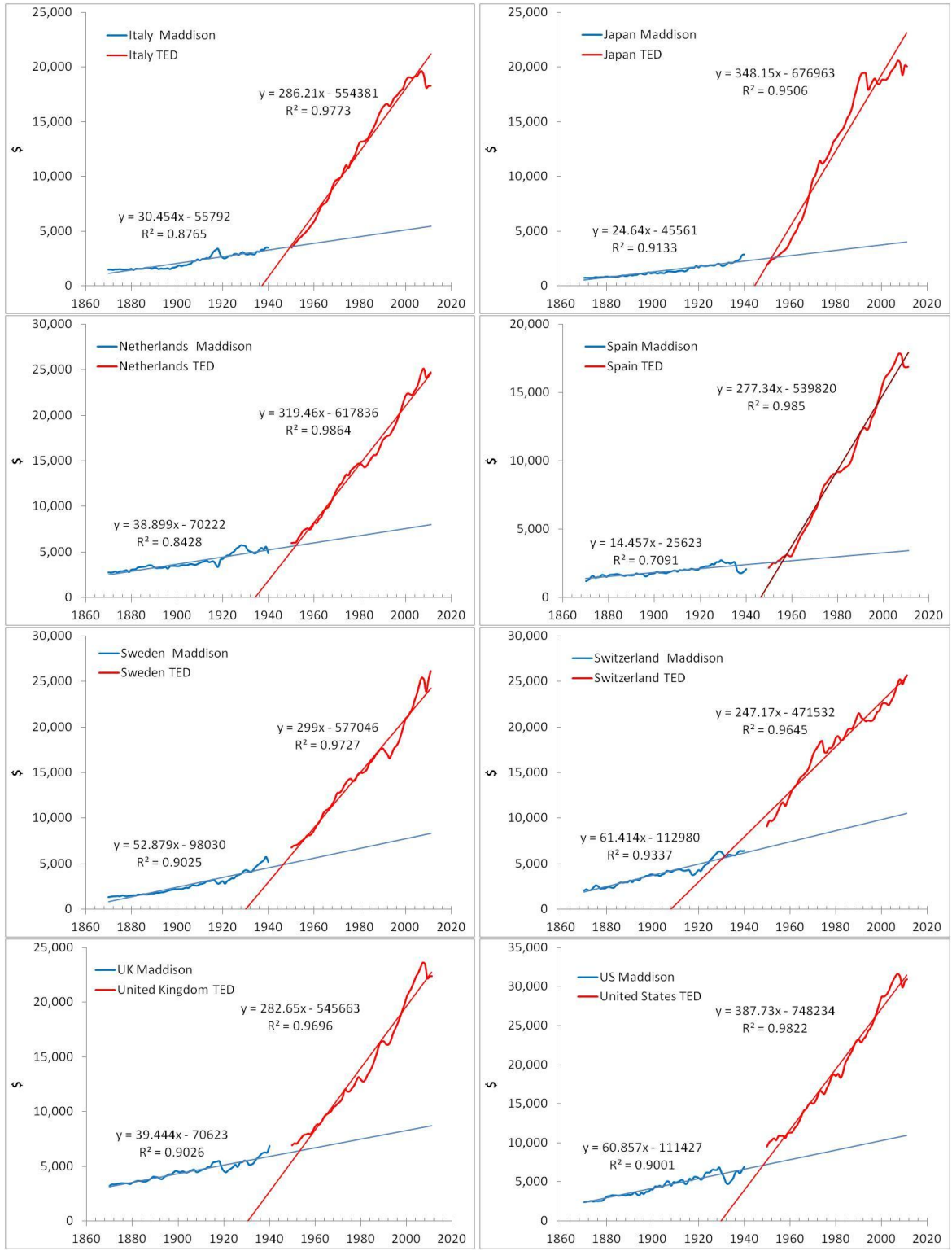

**Figure 2. Real GDP per capita in developed countries from 1870 to 1940 (Maddison historical data) and from 1950 and 2011 (TED). Regression lines are extrapolated into the future and past, respectively.**

The US and UK have been returning to their linear trends during the recent crisis, and thus, both should not wait for any elevated rate of real economic growth. The expectation of a growth rate of 3.5% per year (in terms of real GDP per capita) after 2014 which has been explicitly articulated by the Congressional Budget Office (2012) in its economic outlook is a naïve extrapolation of exponential growth as related to the exponential population growth. The US economy will be rising at a pace dictated by the linear trend.

**Table 3. Real GDP estimates for 2011 from the TED and those extrapolated from the Maddison historical estimates using the relevant trends from 1870 to 1940.**

| Country | TED, $ | Maddison Extrapolated, $ | Ratio |
|---|---|---|---|
| Australia | 25,907 | 7,295 | 3.55 |
| Austria | 24,702 | 5,041 | 4.90 |
| Belgium | 23,999 | 7,240 | 3.31 |
| Canada | 25,297 | 8,421 | 3.00 |
| France | 21,792 | 7,093 | 3.07 |
| Italy | 18,293 | 5,451 | 3.36 |
| Japan | 20,054 | 3,990 | 5.03 |
| Netherlands | 24,712 | 8,004 | 3.09 |
| Spain | 16,874 | 3,450 | 4.89 |
| Sweden | 26,104 | 8,310 | 3.14 |
| Switzerland | 25,640 | 10,524 | 2.44 |
| UK | 22,377 | 8,699 | 2.57 |
| US | 30,928 | 10,956 | 2.82 |

Table 3 lists the current (2011) estimates of real per capita GDP and those obtained by extrapolation of the trends before 1940. The current level is 2.44 (Switzerland) to 5.03 (Japan) higher than the extrapolated one. Therefore, one may consider the current levels as overestimated ones. Then, the relevant GDP deflators have been systematically underestimated since 1950.

## 2. Real GDP, price inflation, and the federal funds rate

Having a new estimate of real GDP per capita in the US one may raise doubts about the accuracy of price inflation estimates. Real GDP is not measured directly but instead defined as the difference between the directly measured nominal GDP and the evaluated GDP deflator. With all measurement problems related to the nominal GDP, this macroeconomic variable is estimated using actual prices. It is likely that some products and services are not counted in (black or grey markets) or double counted. However, the portion of these products and service in the nominal GDP is rather small and likely not changing over time. Therefore, one always has a larger and constant portion of the true nominal GDP which accurately describes the overall evolution of the US economy.

The rate of inflation tells a different story. The diversity of good and services changes at a pace not allowing to directly compare and estimate their relative prices. There are numerous procedures to evaluate qualitatively and then quantitatively the change in consumer prices of practically incomparable goods and services (e.g. film and digital photography) as well as to extrapolate real prices of new products into the past (e.g. computers or mobile phones). With all this turbulent world of prices, the Bureau of Economic Analysis has managed to provide the estimates of GDP deflator at a regular (quarterly and annual) basis.

In Figure 3, we display several curves normalized to their respective levels in 1950: the GDP deflator from 1950 to 2011 as reported by the Bureau of Economic Analysis (green line), the BEA real GDP per capita since 1929 (red line), and the Maddison historical curve from 1870 to 2008 (blue line). Notice that the BEA and Maddison real GDP curves diverge before and coincide after 1950. In order to match the level of 1.0 in 1950, we shifted the Maddison curve between 1870 and 1940 upwards by 0.242 (dashed line). The trend in the shifted Maddison curve is extended through 2011. Then the level of real GDP per capita in 2011 would have been $10,956 instead of $30,928, as estimated by the BEA. This means that the GDP deflator in 2011 was underestimated by a factor of 2.82.

Let's suppose that the real per capita GDP has been evolving along the old trend after 1950. The reported increase in the level of the GDP deflator, dGDP, since 1950 was $dGDP_{2011}/dGDP_{1950}=7.73$. Then we expect that the actual price increase (i.e. reported plus underestimated) was underestimated by (7.73+2.82)/7.73≈1.37 times or by 37%. In other words, the actual rate of price inflation is 37% higher than that reported by the BEA.

This is an interesting and instructive value. It is almost exactly equal to the factor by which the federal funds rate, $R$, exceeds the rate of price inflation as defined by the consumer price index, CPI. (We do not distinguish here the dGDP and CPI.) The interest rate is defined by the Federal Reserve as a major instrument to control price inflation. Figure 4 depicts the headline CPI, the CPI multiplied by a factor of 1.4, and effective rate $R$. The peak values of the scaled CPI and $R$ almost coincide in 1980. Figure 5 displays the corresponding cumulative curves. In the long run, the scaled CPI and $R$ evolve along the same trend and intersect every fifteen to twenty years.

One might assume that the main intention of the FRB is to keep $R$ above the rate consumer price inflation and the higher funds rate should suppress price inflation due to the effect of expensive money. In reality, the FRB has been retaining the interest rate at the long term level of price inflation in order to create neutral conditions for money supply. This could be a wiser prerequisite for a central bank. Actually, the FRB does not need that factor 1.4. The problem is in the wrong estimates of inflation since 1950.

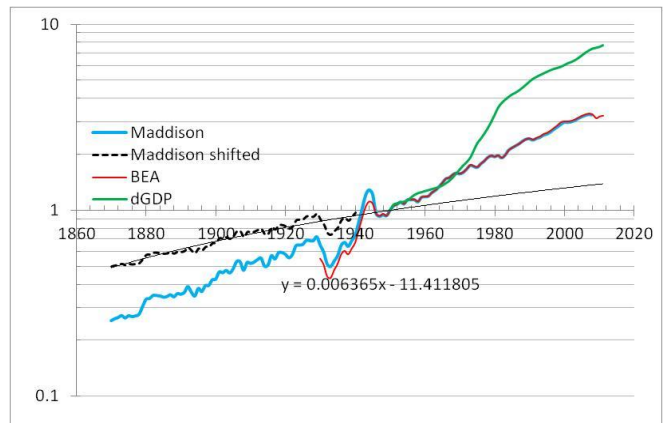

**Figure 3. Real GDP per capita reported by the BEA from 1929 to 2011 and by Maddison from 1870 to 2008. Nominal GDP per capita reported by the BEA. The trend (regression equation in shown in the Figure) of the Maddison time series between 1870 and 1940 is extrapolated in the future. All curves are normalized to 1950. Therefore, the Maddison curve between 1870 and 1940 is shifted up by 0.242 in order to match the normalization constant (1.0) in 1950. Notice the *lin-log* scale.**

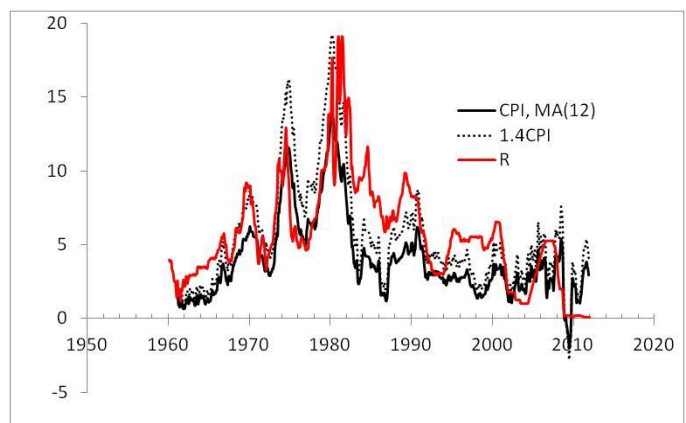

**Figure 4. Monthly estimates of *R*, CPI (MA(12)), and the CPI multiplied by a factor of 1.4.**

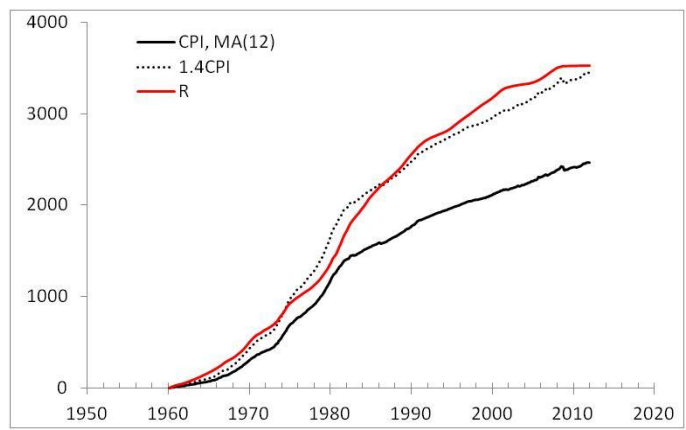

**Figure 5. Cumulative values of the monthly estimates of *R*, CPI, and the CPI multiplied by a factor of 1.4.**

**Conclusion**

We have estimated real GDP per capita in fourteen developed countries and found some new arguments in favor a (piece-wise) constant annual increment since 1870. There is an artificial structural break in all time series between 1940 and 1950 which might be associated with some changes in measuring units/procedures. The latter changes might be expressed in a significant underestimation of the economy-wide price inflation since 1950. Instructively, the degree of underestimation (~37%) is almost exactly equal to the factor by which the federal interest rate exceeds the estimated rate of inflation in the long run. One might assume that the federal rate is actually neutral relative to the actual rate of inflation. Currently, the interest rate is very low because it has to compensate for the large deviation cumulated between 1990 and 2003. The equilibrium (the intercept of the cumulative curves) will be reached in 2013-2015.

**References**

Beechy, M. and P. Ä. Österholm (2008). Revisiting the uncertain unit root in GDP and CPI: testing for non-linear trend reversion, *Economics Letters*, 100, pp. 221-223.

Ben-David, D., R. L. Lumsdaine, and D. H. Papell (2003). Unit roots, postwar slowdowns and long-run growth: Evidence from two structural breaks, *Empirical Economics*, 28, 303.319.

Bureau of Economic Analysis (2011). Concepts and Methods of the U.S. National Income and Product Accounts, http://bea.gov/national/pdf/NIPAchapters1-9.pdf

Bureau of Economic Analysis (2012). Current-dollar and "real" GDP, Table retrieved April 20, 2012 from http://bea.gov/national/index.htm

Caporale, G. M. and L.A. Gil-Alana (2009). Long Memory in US Real Output per Capita, DIW Berlin, May 2009.

Chen, S.-W. (2008). Are 19 Developed Countries' Real Per Capita GDP levels Non-stationary? A Revisit, *Economic Bulletin*, vol. 3, No. 2 pp. 1-11

Conference Board (2012). Total Economy Database, Table retrieved April 25, 2012 from http://www.conference-board.org/data/economydatabase/

Congressional Budget Office (2012). The Budget and Economic Outlook: Fiscal Years 2012 to 2022, http://www.cbo.gov/sites/default/files/cbofiles/attachments/01-31-2012_Outlook.pdf

Cuestas, J. C. and D. Garratt (2008). Is real GDP per capita a stationary process? Smooth transitions, nonlinear trends and unit root testing, Nottingham Trent University Discussion Papers, No. 2008/12

Kitov, I. (2009). The Evolution of Real GDP Per Capita in Developed Countries, Journal of Applied Economic Sciences, Spiru Haret University, Faculty of Financial Management and Accounting Craiova, vol. IV(1(8)_ Summ), pp. 221-234.

Kitov, I. and O. Kitov (2012). Real GDP per capita in developed countries since 1870. MPRA paper 39021, http://mpra.ub.uni-muenchen.de/39021/.

Michelacci, C. and P. Zaffaroni (2000). (Fractional) Beta convergence, *Journal of Monetary Economics* 45, pp. 129-153.

Maddison, A. (2004). Contours of the World Economy and the Art of Macro-measurement 1500-2001, Ruggles Lecture, IARIW 28th General Conference, Cork, Ireland August 2004.

Mayoral, L. (2006). Further evidence on the statistical properties of real GNP, *Oxford Bulletin of Economics and Statistics* 68, pp. 901-920.

Narayan, P.K. (2006). Are G7 real per capita GDP levels non-stationary, 1870.2001? *Japan and the World Economy*, doi:10.1016/j-japwor.2006.08.001

Nelson, C. R. and C. I. Plosser (1982). Trends and random walks in macroeconomic time series: Some evidence and implications, *Journal of Monetary Economics*, 10, 139.162.

Vougas, D. V. (2007). Is the trend in post-WW II US real gdp uncertain or non-linear?, *Economics Letters*, vol. 94, pp. 348-355.